# Noise and dynamics of stimulated Brillouin scattering microresonator lasers

William Loh[*], Scott B. Papp, and Scott A. Diddams

*National Institute of Standards and Technology (NIST), Boulder, CO 80305, USA*

We use theoretical analysis and numerical simulation to investigate the operation of a laser oscillating from gain supplied by stimulated Brillouin scattering (SBS) in a microresonator. The interaction of the forward, backward, and density waves within the microresonator results in a set of coupled-mode equations describing both the laser's phase and amplitude evolution over time. Using this coupled-mode formalism, we investigate the performance of the SBS laser under noise perturbation and identify the fundamental parameters and their optimization to enable low-noise SBS operation. The intrinsic laser linewidth, which is primarily limited by incoherent thermal occupation of the density wave, can be of order hertz or below. Our analysis also determines the SBS laser's relaxation oscillation, which results from the coupling between the optical and density waves, and appears as a resonance in both the phase and amplitude quadratures. We further explore contributions of the pump noise to the SBS laser's performance, which we find under most circumstances to increase the SBS laser noise beyond its fundamental limits. By tightly stabilizing the pump laser onto the microcavity resonance, the transfer of pump noise is significantly reduced. Our analysis is both supported and extended through numerical simulations of the SBS laser.

## I. INTRODUCTION

Narrow linewidth lasers serve as the essential enabler for a variety of applications in high-precision spectroscopy [1], remote sensing [2], low-noise microwave signal generation [3], and coherent optical communications [4]. One common technique used to produce narrow linewidths is through stabilization of a seed laser to a high-Q cavity. This technique is capable of reducing the laser linewidth below 1 Hz but comes at the added cost of system complexity. Furthermore, the locking bandwidth that is typically achievable is < 1 MHz and thus prevents stabilization of the laser noise beyond this range. For many applications, an intrinsically low-noise laser source is often desirable.

Lasers with intrinsically narrow linewidth are commonly achieved through an external cavity configuration [5–12]. The additional passive cavity length increases the density of resonant modes within the laser's gain bandwidth and thus reduces the amount of noise power coupled into each mode [13, 14]. By taking advantage of this property, semiconductor external cavity lasers operating at 1550 nm wavelength typically achieve linewidths in the range of 10 kHz to 100 kHz [9, 15–17]. Although fiber lasers exhibit linewidths below that of their semiconductor counterparts [18], their larger footprint makes them incompatible with a full chip-integrated solution.

While an external cavity configuration is beneficial for reducing laser noise, the cavity length cannot be extended indefinitely due to the eventual appearance of sidemodes that cannot be rejected by an intracavity filter. Another possible route towards producing narrow linewidth lasers is to reduce the intracavity losses, thereby clamping the noise at a lower lasing threshold, while still maintaining a high-power oscillation signal. This strategy generally applies to all lasers but is especially effective in lasers which rely on the nonlinear stimulated Brillouin scattering (SBS) process [19] for generating the gain necessary for self-oscillation [20–22]. The SBS gain is unique as its gain bandwidth is only ~10 to 100 MHz for typical dielectric materials used in confining light, which can be many orders of magnitude narrower than conventional intracavity grating filters. This narrow gain bandwidth enables single-mode operation for cavity lengths on the order of 10 m. Since the cavity mode spacing is ideally designed to be larger than the gain bandwidth, approximately all of the noise within the 10 to 100 MHz span couples into the oscillating cavity mode [13, 23, 24].

Recently, the potential for low-noise SBS oscillation was predicted and demonstrated in high-Q $CaF_2$ [25] and silica microresonators [26–29] and also in highly-nonlinear chalcogenide waveguides [30]. Unlike conventional SBS lasers requiring ~10 m of cavity length to yield enough SBS gain to compensate for losses [22], oscillation was made possible in [25–29] due to the low intrinsic losses (high Q) of the microresonator cavity. These low losses reduce the requirements for the SBS gain at threshold, further clamping the level of injected SBS noise. To generate significant SBS gain, the Stokes wave must be precisely matched to fall within the gain bandwidth of the nonlinear interaction. This condition is readily achieved in microresonators through the ability to accurately control cavity dimensions. These properties make the SBS microresonator laser an ideal choice for a compact, narrow linewidth laser source.

The goal of this work is to formulate a description of the SBS laser through a set of coupled-mode equations for the forward, backward, and density waves. To our knowledge, no comprehensive investigation of the SBS laser exists for either bulk or microcavity operation. Although our analysis will be specifically directed to the case of a microresonator SBS laser, our obtained results can be extended to any laser operating via SBS gain. We use the developed coupled-mode equations to analyze the steady-state operation of the SBS laser and to determine the response of the laser to small-signal noise perturbations.

Conventionally, the SBS laser noise is treated either by deriving an equivalent Schawlow-Townes linewidth relation to account for SBS lasing [25, 28] or by analyzing the SBS laser from the perspective of a filter that acts to reduce pump fluctuations [22, 31]. Building on the previous work, we solve the noise of the SBS laser by treating the laser's response to perturbations of the density wave. In contrast to traditional lasers, the frequencies of the pump, SBS, and density waves are all correlated with one another which results in feedback when one of the waves is perturbed. This feedback modifies the fundamental noise limit of the laser and also induces the transfer of pump noise into the SBS wave [25], an effect that gives the pump the appearance of being filtered. In this regard, we view the SBS laser to be similar to the Erbium fiber laser in which the intrinsic noise limits of the SBS laser can be reached only if the transfer of pump noise to the oscillation signal can be made small.

In the second half of our work, we develop our noise model of the SBS laser investigating the laser's intrinsic noise limits and also the conversion of pump noise into SBS noise. We find that the combination of a high-Q cavity along with a nonlinear noise process that preferentially favors high optical powers allows the SBS laser's intrinsic noise to be significantly lower than that of typical lasers. Furthermore, we show that the coupling between waves results in a relaxation oscillation resonance in both amplitude and phase that acts to damp noise fluctuations at high frequencies. We conclude with numerical simulations of the SBS laser, which serve to both support our analysis and to highlight the fundamental noise performance of the system. In particular, we show that the ratio of noise to signal in SBS lasers allows for oscillation linewidths in the range of hertz or below.

## II. SBS RESULTS SUMMARY

In Sections III to VI, we present a detailed analysis of both the SBS laser's operation in steady state and also the laser's response to noise. Since the main results of our analysis can often be lost within our derivations, we use this section to highlight the central equations of our work.

In Section IV, Eqs. (14) and (15) summarize the coupled-mode equations governing the (noiseless) interaction of the forward propagating, SBS, and density waves. To the level of the physics captured, the entirety of the SBS laser's operation can be solved using either Eq. (14) or (15). Thus these equations serve as a useful starting point for numerical simulations or detailed analysis.

Equations (19) and (22) describe the steady-state operation of the SBS laser. In particular, Eq. (19) quantifies the amplitudes of the forward and SBS waves as a function of the laser's bias point. Under steady-state conditions, the forward wave becomes clamped, and the excess pump power is used to fuel the SBS oscillation. Equation (22) solves for the fundamental input-output relation that governs the behavior of every oscillator system. The SBS output power varies linearly with the power coupled into the microresonator with a characteristic slope and threshold that can be determined from Eq. (22).

In Section V, we investigate the amplitude noise induced on the SBS laser through (1) noise fluctuations of the pump and (2) thermally-excited fluctuations of the density wave. The pump noise case allows us to analyze the transfer of pump noise into the SBS wave, while our treatment of phonon perturbations allows us to quantify the SBS laser's fundamental noise limit. To make our analysis tractable, we assume ideal (zero pump and SBS gain detuning) operating conditions. Equations (28) and (30) describe the spectral density of amplitude fluctuations for pump and phonon perturbations, respectively. The example in Section VIII shows how to relate these spectral densities to laser relative intensity noise (RIN). Equation (26) approximates the relaxation oscillation frequency of the SBS laser (valid for high-Q cavities), which occurs due to the laser's inherent amplitude feedback. Equation (27) approximates the damping of this relaxation oscillation.

Section VI analyzes the fundamental limit of the SBS laser's frequency noise under zero detuning conditions. If necessary, frequency noise can be readily converted to phase noise or power spectral density using conventional methods [32]. The spectral density governing frequency fluctuations is solved in Eq. (34), while the corresponding white-noise floor is identified in Eq. (35). Equation (37) presents an approximation to the white frequency noise level for the case of a high-Q cavity. To reduce noise, the lasing threshold should be minimized thereby clamping the level of injected noise, while the SBS power should be simultaneously maximized. The SBS laser's low intracavity losses and high circulating optical powers are well-suited to the purpose of maximizing the ratio of signal to noise. Finally, Eqs. (38) and (39) quantify the SBS laser's resonance frequency and damping due to phase feedback.

## III. SBS COUPLED-MODE EQUATIONS

In this section, we use the coupled-mode formalism to describe the SBS interaction between the forward, backward, and density waves for a microresonator oscillator. The derivation of the coupled-mode equations is based on standard treatments of the nonlinear SBS process, but modified to account for the physics of the microresonator cavity. We consider here a general spherical microresonator configuration with radius $R$; however, this analysis can be readily extended to other cavity geometries. A schematic of our system is shown in Fig. 1 consisting of a microresonator pumped by a continuous wave (CW) laser and generating a counterpropagating SBS wave. We begin our analysis of this system using the traditional electromagnetic wave equation [33]

$$\left(\nabla^2 - \frac{n^2}{c^2}\frac{\partial^2}{\partial t^2}\right)\tilde{E}_{B,F} = \frac{1}{\epsilon_0 c^2}\frac{\partial^2 \tilde{P}_{B,F}}{\partial t^2} \quad (1)$$

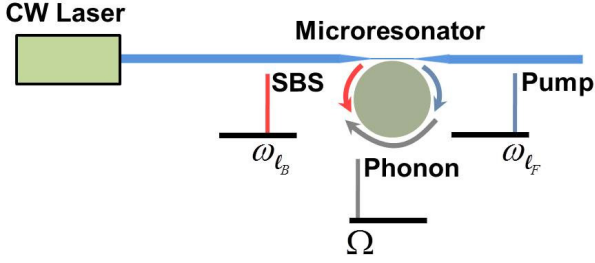

Fig. 1. Schematic of setup for SBS generation via a microresonator oscillator. The CW laser pumps the microresonator to generate SBS oscillation in the reverse direction.

with [34, 35]

$$\tilde{E}_F = \frac{1}{2} A_F(t) e^{i(\omega_{\ell_F} t - \ell_F \phi)} F_F(r,\theta) + \text{c.c.}$$
$$\tilde{E}_B = \frac{1}{2} A_B(t) e^{i(\omega_{\ell_B} t + \ell_B \phi)} F_B(r,\theta) + \text{c.c.} \quad (2)$$

denoting the individual forward ($F$) and backward ($B$) propagating electric fields and their complex conjugates (c.c.). In Eq. (1), $n$ is the refractive index of the medium, $c$ is the speed of light, and $\epsilon_0$ is the permittivity of free space. $A_F, \omega_{\ell_F}, \ell_F, F_F$ ($A_B, \omega_{\ell_B}, \ell_B, F_B$) represent the amplitude, frequency, angular momentum, and optical mode profile of the forward (backward) propagating field. The coordinates $r, \theta, \phi$ denote the radial distance, zenith angle, and azimuthal angle of the microresonator system.

Because of loss in the microresonator, the refractive index of Eq. (1) is composed of both real and imaginary contributions. We may thus represent this index as $n = n_0 - i n_{loss}$ where $n_0$ is the real component of the index and $n_{loss}$ is the imaginary component accounting for both intrinsic material losses and external coupling loss [36]. Although strictly speaking the external coupling loss is defined only along discrete points of the cavity where the field is coupled to an external wave, we may define an effective $n_{loss}$ which averages the total coupling loss over the period of one round-trip. If we further assume the microresonator losses to be small (high Q cavity), we may treat the imaginary index as a perturbation to the overall index with the result that

$$n^2 \approx n_0^2 - 2 i n_0 n_{loss} \quad (3)$$

In Eq. (1), the nonlinear polarization $\tilde{P}$ generated through the interaction of the electric field with a propagating density wave $\tilde{\rho}$ can be described through [33]

$$\tilde{P} = \epsilon_0 \rho_0^{-1} \gamma_e \tilde{\rho} \tilde{E} \quad (4)$$

where $\rho_0$ is the equilibrium density of the material and $\gamma_e$ is the electrostrictive constant. For the SBS process, this polarization describes the scattering of the forward wave off of the density wave, which provides amplification for the backward wave. The density wave can be expressed as

$$\tilde{\rho} = \rho_0 + \frac{1}{2} \rho(t) e^{i(\Omega t - \ell_\rho \phi)} F_\rho(r,\theta) + \text{c.c.} \quad (5)$$

where $\rho(t), \Omega, \ell_\rho, F_\rho$ denote the amplitude, frequency, angular momentum, and mode profile of the density wave. Substitution of Eqs. (2) and (5) into Eq. (4) yields

$$\tilde{P}_F = \frac{1}{4} \epsilon_0 \rho_0^{-1} \gamma_e \rho A_B e^{i(\omega_{\ell_F} t - \ell_F \phi)} F_\rho(r,\theta) F_B(r,\theta) + \text{c.c}$$
$$\tilde{P}_B = \frac{1}{4} \epsilon_0 \rho_0^{-1} \gamma_e \rho^* A_F e^{i(\omega_{\ell_B} t - \ell_B \phi)} F_\rho^*(r,\theta) F_F(r,\theta) + \text{c.c} \quad (6)$$

Note that we have used $\ell_F + \ell_B = \ell_\rho$ and $\omega_{\ell_F} = \omega_{\ell_B} + \Omega$ in Eq. (6), which together govern the phase-matching requirements for the angular momentum and frequency of the forward, backward, and density waves. Because of the periodic boundary conditions inherent in the system, there is no guarantee that the resulting optical and acoustic modes of the microresonator simultaneously satisfy both conditions for phase-match. For these cases, we assume the acoustic wave acquires the necessary $\ell_\rho, \Omega$ to satisfy phase match. The incurred phase rotation is then accounted for in the density wave equation.

Finally, using Eqs. (2), (3), and (6) in Eq. (1) and grouping together phase-matched terms, we find a coupled set of equations for the forward and backward wave. These coupled-mode equations can be expressed as

$$\frac{\partial A_F}{\partial t} = -\frac{1}{2\tau_F} A_F - i \frac{\gamma_e \omega_{\ell_F}}{4 n_0^2 \rho_0} \rho A_B \Lambda_F + \sqrt{\frac{1}{\tau_{ext}}} S e^{i(\omega_S - \omega_{\ell_F})t}$$
$$\frac{\partial A_B}{\partial t} = -\frac{1}{2\tau_B} A_B - i \frac{\gamma_e \omega_{\ell_B}}{4 n_0^2 \rho_0} \rho^* A_F \Lambda_B \quad (7)$$

with [34]

$$\Lambda_F = \frac{\int_A F_\rho(r,\theta) F_B(r,\theta) F_F^*(r,\theta) dA}{\int_A F_F(r,\theta) F_F^*(r,\theta) dA}$$
$$\Lambda_B = \frac{\int_A F_\rho^*(r,\theta) F_F(r,\theta) F_B^*(r,\theta) dA}{\int_A F_B(r,\theta) F_B^*(r,\theta) dA} \quad (8)$$

In the derivation of Eq. (7), we have used the defining equation for the optical mode in the cavity $\left( \nabla^2 + n^2 \omega_{B,F}^2 / c^2 \right) \left[ F_{B,F} e^{-i\ell_{B,F} \phi} \right] = 0$ along with the slowly-varying approximation for the optical fields. Equation (8) expresses the mode overlaps of the forward, backward, and density waves and thus quantifies their coupling strength. For most microresonators, the forward and backward waves are sufficiently close in angular mode number that their mode profiles do not differ significantly.

In Eq. (8), $\tau_F$ and $\tau_B$ represent the lifetimes for the forward and backward waves. Their inverses physically correspond to the linewidths of the associated optical modes. They are related to $n_{loss}$ through

$$\frac{1}{\tau_{B,F}} = \frac{2\omega_{\ell_{B,F}} \int_A n_{loss_{B,F}} F_{B,F}(r,\theta) F^*_{B,F}(r,\theta) dA}{n_0 \int_A F_{B,F}(r,\theta) F^*_{B,F}(r,\theta) dA} \quad (9)$$

The integral in Eq. (9) represents the confinement factor of the optical mode to the region of loss. Note that in Eqs. (7) and (9), we have allowed these loss rates to be different between the two modes. We have also added a phenomenological pump parameter [37] to Eq. (7) (last term on the right-hand side) with $S, \omega_S$ denoting the amplitude and frequency of the pump field and $1/\tau_{ext}$ denoting the external coupling rate of the pump field into the microresonator. The detuning of the pump from the cavity resonance is quantified by $\omega_S - \omega_{\ell_F}$. Note that in the notation of Eq. (7), $|A_{B,F}|^2$ is proportional to energy, while $|S|^2$ is proportional to power. The detailed analysis of this pump parameter can be found in Ref. [37].

Equation (7) shows that the evolution of the forward (backward) wave is governed by the interaction of the backward (forward) wave and the density wave. As we will see later, the phase of the interaction is such that the forward wave experiences attenuation supplying power for the growth of the backward wave. To complete the description of Eq. (5), we also need to specify the interaction of the optical waves, which reinforces the generation of the density wave via the electrostriction process. We assume the material density satisfies the acoustic wave equation [33]

$$\frac{\partial^2 \tilde{\rho}}{\partial t^2} - \Gamma' \nabla^2 \frac{\partial \tilde{\rho}}{\partial t} - v^2 \nabla^2 \tilde{\rho} = -\frac{1}{2} \epsilon_0 \gamma_e \nabla^2 \langle \tilde{E}^2 \rangle \quad (10)$$

where $\Gamma'$ is a parameter specifying the damping of the density wave and $v$ is the velocity of the wave. Next, we introduce Eqs. (2) and (5) into Eq. (10) and assume slowly varying field amplitudes while grouping together phase-matched terms. From this procedure, we determine the evolution of the density wave to obey

$$\frac{\partial \rho}{\partial t} = i\frac{\Omega_b^2 - \Omega^2}{2\Omega} \rho - \frac{\Gamma_b}{2} \rho - i\frac{\epsilon_0 \gamma_e}{4\Omega} \frac{\ell_\rho^2}{R^2} A_F A_B^* \Lambda_\rho \quad (11)$$

with

$$\Lambda_\rho = \frac{\int_A F_F(r,\theta) F_B^*(r,\theta) F_\rho^*(r,\theta) dA}{\int_A F_\rho(r,\theta) F_\rho^*(r,\theta) dA} \quad (12)$$

representing the overlap of the forward, backward, and density waves. In Eq. (11), $R$ denotes the radius of the microresonator cavity, $\Gamma_b = \Gamma' \Omega_b^2 / v^2$ denotes the loss rate of the density wave, and $\Omega_b$ denotes the frequency where the SBS gain is maximum as determined from the definition of the acoustic mode $(\nabla^2 + \Omega_b^2/v^2)[F_\rho e^{-i\ell_\rho \phi}] = 0$. Note that in defining $\Gamma_B$, we have assumed $\Gamma'$ to be approximately constant over the mode area of the density wave. If necessary, one can readily define an effective confinement factor as was done for the optical losses in Eq. (9).

In Eq. (11), we have also assumed that the mode profile of the forward and backward fields to be that of the fundamental transverse mode and that its spatial derivatives are predominantly in the azimuthal direction. The cavity mode is located near the boundaries of the resonator centered on the equator, and thus we approximate $(r \approx R, \theta \approx \pi/2)$ [36]. These approximations can be relaxed given knowledge of the exact mode distribution. Note that the first term on the right-hand side of Eq. (11) accounts for the phase rotation induced by mismatch between $\Omega$ and the frequency of the acoustic mode $(\Omega_b)$.

### IV. SBS LASER STEADY-STATE OPERATION

Together, Eqs. (7) and (11) provide the set of coupled-mode equations describing the SBS generation process in a microresonator oscillator. The forward wave supplies the power necessary for amplification of the backward wave and is replenished by the external pump to maintain steady-state operation. If the gain of the backward and density waves exceed their respective losses, a large-signal oscillation can develop. To proceed further with our analysis, we first note that Eq. (7) in its current form does not yield a steady-state solution for $A_F$ if we assume $A_F, A_B, \rho$ to be stationary with time. This can readily be seen if we set $\partial A_F / \partial t = 0$ in Eq. (7) and then attempt to solve the resulting equation. The problem results from the frequency detuning of the pump from the cavity resonance and can be remedied if we define

$$\begin{aligned} A'_F &= A_F e^{-i\sigma_F t} \\ A'_B &= A_F e^{-i\sigma_B t} \\ \rho' &= \rho e^{-i\sigma_\rho t} \end{aligned} \quad (13)$$

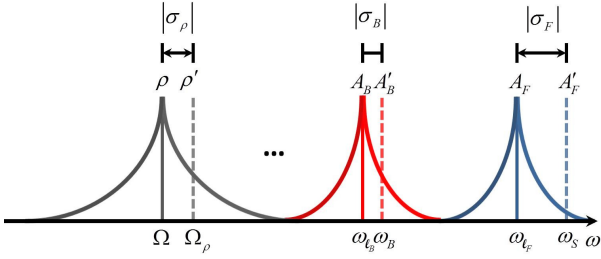

Fig. 2. Illustration of forward, backward, and density wave translations due to pump detuning. Note that although the shifts are in general different for each of the waves, the combined shift must satisfy $\sigma_F = \sigma_B + \sigma_\rho$.

where $\sigma_F = \omega_S - \omega_{\ell_F}$, $\sigma_B = \omega_B - \omega_{\ell_B}$, and $\sigma_\rho = \Omega_\rho - \Omega$. Here, $\omega_B$ ($\Omega_\rho$) is the final frequency of the backward wave (density wave) translated from its original frequency at $\omega_{\ell_B}$ ($\Omega$). The frequency translations are such that sum of the backward and density wave translations is equal to the translation of the forward wave (or $\sigma_F = \sigma_B + \sigma_\rho$), thereby maintaining the phase-matching condition between all three waves. Note that unlike the case of a continuous waveguide, the boundary conditions of the resonator necessitate the existence of discrete modes. Thus, the detuning of the pump forces a shift in both the SBS and density waves to preserve phase-match. These frequency translations can be visualized through Fig. 2 [38]. Note that since both optical and acoustic modes are plotted in Fig. 2, the vertical axis of the figure does not strictly correspond to any physical quantity. Nevertheless, the horizontal axis depicts the frequency translations of the forward, backward, and density waves that must occur to compensate for the pump detuning.

The redefinition of Eq. (13) transforms Eqs. (7) and (11) into

$$\frac{\partial A'_F}{\partial t} = -\frac{A'_F}{2\tau_F} - i\frac{\gamma_e \omega_{\ell_F}}{4n_0^2 \rho_0}\rho' A'_B \Lambda_F - i\sigma_F A'_F + \sqrt{\frac{1}{\tau_{ext}}}S$$

$$\frac{\partial A'_B}{\partial t} = -\frac{A'_B}{2\tau_B} - i\frac{\gamma_e \omega_{\ell_B}}{4n_0^2 \rho_0}\rho'^* A'_F \Lambda_B - i\sigma_B A'_B \quad (14)$$

$$\frac{\partial \rho'}{\partial t} = i\frac{\Omega_b^2 - \Omega^2}{2\Omega}\rho' - \frac{\Gamma_b}{2}\rho' - i\frac{\epsilon_0 \gamma_e}{4\Omega}\frac{\ell_\rho^2}{R^2}A'_F A'^*_B \Lambda_\rho - i\sigma_\rho \rho'$$

It is clear that the time dependence of Eq. (7) has been effectively removed in Eq. (14). The steady-state solutions of the coupled-mode equations are thus shifted versions of the cavity modes with the forward wave translated to the frequency of the pump and with the backward and density waves translated to some intermediate frequency. We now determine these frequency shifts by attempting to find the steady-state solution of Eq. (14). In order to simplify our analysis, we assume $\omega_\ell = \omega_{\ell_F} \approx \omega_{\ell_B}$ since $\Omega \ll \omega_{\ell_F}, \omega_{\ell_B}$. We begin by using $A'_F = |A'_F|e^{i\phi_F}$, $A'_B = |A'_B|e^{i\phi_B}$, $\rho' = |\rho'|e^{i\phi_\rho}$, $S = |S|e^{i\phi_S}$ to separate Eq. (14) into its individual coupled-mode equations for amplitude and phase

$$\frac{\partial |A'_F|}{\partial t} = -\frac{|A'_F|}{2\tau_F} + \frac{\gamma_e \omega_\ell}{4n_0^2 \rho_0}|\rho'||A'_B|\Lambda_F \sin(\phi_\rho + \phi_B - \phi_F)$$
$$+ \sqrt{\frac{1}{\tau_{ext}}}|S|\cos(\phi_S - \phi_F)$$

$$\frac{\partial \phi_F}{\partial t} = -\sigma_F - \frac{\gamma_e \omega_\ell}{4n_0^2 \rho_0}\frac{|\rho'||A'_B|}{|A'_F|}\Lambda_F \cos(\phi_\rho + \phi_B - \phi_F)$$
$$+ \sqrt{\frac{1}{\tau_{ext}}}\frac{|S|}{|A'_F|}\sin(\phi_S - \phi_F)$$

$$\frac{\partial |A'_B|}{\partial t} = -\frac{|A'_B|}{2\tau_B} + \frac{\gamma_e \omega_\ell}{4n_0^2 \rho_0}|\rho'||A'_F|\Lambda_B \sin(\phi_F - \phi_B - \phi_\rho) \quad (15)$$

$$\frac{\partial \phi_B}{\partial t} = -\sigma_B - \frac{\gamma_e \omega_\ell}{4n_0^2 \rho_0}\frac{|\rho'||A'_F|}{|A'_B|}\Lambda_B \cos(\phi_F - \phi_B - \phi_\rho)$$

$$\frac{\partial |\rho'|}{\partial t} = -\frac{\Gamma_b}{2}|\rho'| + \frac{\epsilon_0 \gamma_e \ell_\rho^2}{4\Omega R^2}|A'_F||A'_B|\Lambda_\rho \sin(\phi_F - \phi_B - \phi_\rho)$$

$$\frac{\partial \phi_\rho}{\partial t} = -\sigma_\rho + \frac{\Omega_b^2 - \Omega^2}{2\Omega} - \frac{\epsilon_0 \gamma_e \ell_\rho^2}{4\Omega R^2}\frac{|A'_F||A'_B|}{|\rho'|}\Lambda_\rho \cos(\phi_F - \phi_B - \phi_\rho)$$

Here, $\phi_F, \phi_B, \phi_\rho, \phi_S$ denote the phases of the forward, backward, density, and pump fields. It is important to note that Eq. (15) assumes $\Lambda_F, \Lambda_B, \Lambda_\rho$ to have zero phase, which is generally true only if the mode overlaps and thus $F_F, F_B, F_\rho$ of Eqs. (8) and (12) have zero phase. Since the forward, backward, and density waves are all represented by a complex amplitude multiplied by an eigenmode [see Eqs. (2) and (5)], one can in most cases reassign the eigenmode phase to the phase of the field amplitude.

In the steady state, all time derivatives of Eq. (15) must independently yield zero. The simultaneous solution of the $\partial |A'_B|/\partial t$ and $\partial |\rho'|/\partial t$ amplitude equations yields a relation between $|A'_B|^2$ and $|\rho'|^2$. The simultaneous solution of the $\partial \phi_B/\partial t$ and $\partial \phi_\rho/\partial t$ phase equations yields a second relation between $|A'_B|^2$ and $|\rho'|^2$. The joint solution of these relations along with $\sigma_F = \sigma_B + \sigma_\rho$ yields

$$\sigma_B = \frac{\sigma_F - \left[(\Omega_b^2 - \Omega^2)/2\Omega\right]}{1 + \Gamma_b \tau_B}$$

$$\sigma_\rho = \frac{\sigma_F \Gamma_b \tau_B + \left[(\Omega_b^2 - \Omega^2)/2\Omega\right]}{1 + \Gamma_b \tau_B} \quad (16)$$

Note that the frequency shifts are asymmetric for the backward and density waves. For high-Q cavities, the SBS loss is much lower than the density wave loss ($\Gamma_\rho \tau_B \gg 1$) and thus the frequency translation is larger for the density wave. Equation (16) has the interpretation that because of the pump detuning, the forward wave must shift from its natural cavity resonance to the pump frequency by the amount $\sigma_F$. If the frequency of the SBS wave remains unaltered in this process, then the density wave must absorb the entirety of the frequency shift to preserve phase match. This is a valid steady-state solution $(\sigma_B = 0)$ as long as the frequency of the density wave shifts so that it exactly falls on the frequency where the SBS gain is maximum $[\sigma_\rho = \sigma_F = (\Omega_b^2 - \Omega^2)/2\Omega]$. If the density wave does not end up at the peak of the SBS gain, then a residual phase shift exists for $\partial \varphi_\rho / \partial t$ in Eq. (15) that prevents $\partial \varphi_B / \partial t$ from simultaneously reaching steady state. Thus, in the general case, both the steady-state backward and density waves experience a translation in frequency due to the pump detuning. Note that the residual phase rotation of the SBS gain exists in $\partial \varphi_\rho / \partial t$ even without a pump detuning. In these cases, the backward and density waves are "pulled" towards the gain maximum.

The nature of the SBS gain can be made apparent if we solve for $|\rho'|$ in Eq. (15) using $\partial |\rho'| / \partial t = 0$. This analysis yields

$$|\rho'| = \frac{\epsilon_0 \gamma_e \ell_\rho^2}{2\Omega \Gamma_b R^2} |A_F'||A_B'| \Lambda_\rho \sin(\phi_F - \phi_B - \phi_\rho) \quad (17)$$

Substitution of Eq. (17) into the forward and backward wave amplitude equations yields

$$\frac{\partial |A_F'|}{\partial t} = -\frac{|A_F'|}{2\tau_F} - \frac{\epsilon_0 \gamma_e^2 \omega_\ell \ell_\rho^2 |A_F'||A_B'|^2}{8 n_0^2 \rho_0 \Gamma_b \Omega R^2} \Lambda_F \Lambda_\rho \sin^2(\phi_F - \phi_B - \phi_\rho)$$
$$+ \sqrt{\frac{1}{\tau_{ext}}} |S| \cos(\phi_S - \phi_F)$$
$$\frac{\partial |A_B'|}{\partial t} = -\frac{|A_B'|}{2\tau_B} + \frac{\epsilon_0 \gamma_e^2 \omega_\ell \ell_\rho^2 |A_F'|^2 |A_B'|}{8 n_0^2 \rho_0 \Gamma_b \Omega R^2} \Lambda_B \Lambda_\rho \sin^2(\phi_F - \phi_B - \phi_\rho)$$
(18)

In Eq. (18), the first term of each equation corresponds to the loss of the field amplitude over time due to intrinsic material losses or external coupling loss. The second term corresponds to attenuation for the forward wave and to SBS gain for the backward wave. Note that the term $\sin^2(\phi_F - \phi_B - \phi_\rho)$ is always non-negative, and therefore the phases of the fields in the SBS process are such that the forward wave always supplies power for the amplification of the backward wave. If we further assume that the pump detuning exactly cancels the SBS gain detuning $[\sigma_F = (\Omega_b^2 - \Omega^2)/2\Omega]$ such that the backward wave is operated at the SBS gain maximum, we find from Eq. (16) that $\sigma_B = 0$ and $\sigma_\rho = \sigma_F$. This then implies $\phi_F - \phi_B - \phi_\rho = \pi/2$ to satisfy the steady state of Eq. (15). Thus, these conditions yield the phase arrangement required of the individual waves for the largest SBS gain. However, since the SBS gain is also proportional to $|A_F'|^2$, the maximum gain is achieved only when the pump detuning is also zero $(\sigma_F = 0)$.

The steady-state solution of Eq. (18) yields the amplitudes of the forward and backward waves depending on the operating parameters of the microresonator laser. With $\partial |A_B'|/\partial t = 0, \partial |A_F'|/\partial t = 0$ and assuming $\Lambda_F \approx \Lambda_B$, we find that

$$|A_F'|^2 = \frac{4 n_0^2 \rho_0 \Gamma_b \Omega R^2}{\tau_B \epsilon_0 \gamma_e^2 \omega_\ell \ell_\rho^2 \Lambda_B \Lambda_\rho} \frac{1}{\sin^2(\phi_F - \phi_B - \phi_\rho)}$$
$$|A_B'|^2 = \frac{2\tau_B}{\sqrt{\tau_{ext}}} |S||A_F'| \cos(\phi_S - \phi_F) - \frac{\tau_B}{\tau_F} |A_F'|^2$$
(19)

Equation (19) shows that the amplitude of the forward wave is constant once the phase relationship of the forward, backward, and density waves is known. As before, this phase relationship is dependent on the pump detuning and SBS gain detuning of the microresonator. The pump power clamps once the SBS threshold is reached since the SBS gain must saturate to the level of the total cavity loss. This gain saturation necessitates the saturation of $|A_F'|^2$, since it is the forward wave power that drives the SBS gain of the microresonator. The dependence of $|A_F'|^2$ on $\sin^2(\phi_F - \phi_B - \phi_\rho)$ is a result of the dependence of the SBS gain on this phase relationship [see Eq. (18)]. The SBS gain is optimally phase-matched when $\phi_F - \phi_B - \phi_\rho = \pi/2$, and thus less forward wave power is required to compensate for system loss.

In Eq. (19), the expression for $|A_B'|^2$ can be understood through a rearrangement by first dividing by $\tau_B$. Next, we separate $|A_F'|^2/\tau_F$ into its two components $|A_F'|^2/\tau_{0,F}$ and $|A_F'|^2/\tau_{ext}$, which describe the intrinsic material losses $(\tau_{0,F})$ and coupling losses $(\tau_{ext})$ of the forward wave. Moving $|A_F'|^2/\tau_{0,F}$ to the left-hand side, we find

$$\frac{|A'_B|^2}{\tau_B} + \frac{|A'_F|^2}{\tau_{0,F}} = \frac{2}{\sqrt{\tau_{ext}}}|S||A'_F|\cos(\phi_S - \phi_F) - \frac{|A'_F|^2}{\tau_{ext}} \quad (20)$$

The left side of Eq. (20) describes the flow rate of energy out of the microresonator, whereas the right-hand side describes the net rate of energy flow into the system [37]. Thus Eq. (20) is a statement of power conservation. The form of the right-hand side of Eq. (20) is deceiving as it would appear that $|A'_B|^2$ is related to a cross-term between the pump and forward waves. This is remedied if we use the identity $2|S||A'_F|\cos(\phi_S - \phi_F) = SA'^*_F + S^*A'_F$ in Eq. (20) and set $A'_F = (T+S)\sqrt{\tau_{ext}}$ with $T$ representing the transmitted wave past the microresonator. The detailed derivation for this form of $A'_F$ can be found in Refs. [37, 39]. However, we note that through division by $\sqrt{\tau_{ext}}$ and rearrangement of $S$ onto the left-hand side, we find the condition necessary for the cancellation of the transmitted pump field via the field leaking out of the cavity. Upon substitution of $A'_F$ into Eq. (20), we find

$$\frac{|A'_B|^2}{\tau_B} + \frac{|A'_F|^2}{\tau_{0,F}} = |S|^2 - |T|^2 \quad (21)$$

Therefore, we see that the energy and thus power of the backward wave is proportional to the net power that is coupled into the cavity minus that which is dissipated in the forward wave. Note that Eq. (21) can be rewritten as

$$\frac{|A'_B|^2}{\tau_{ext}} = \frac{\tau_B}{\tau_{ext}}\left[\left(|S|^2 - |T|^2\right) - \frac{|A'_F|^2}{\tau_{0,F}}\right] \quad (22)$$

which shows that the SBS power coupled out has the traditional form of a laser's input-output relation with $|A'_F|^2/\tau_{0,F}$ serving as the threshold optical power and $\tau_B/\tau_{ext}$ serving as the slope efficiency.

**V. SBS LASER AMPLITUDE SMALL-SIGNAL ANALYSIS**

Here, we use the coupled-mode equations of Eq. (15) to analyze the response of the SBS oscillator to small-signal perturbations. Without any assumptions on operating parameters, this task is difficult as it requires the simultaneous solution of six coupled equations. For example, as can be seen from Eq. (15), if one were to vary the detuning of the pump from the cavity resonance $(\sigma_F)$, the condition $\partial\phi_F/\partial t = 0$ forces the phases of the individual waves to rotate in order to satisfy steady state. This phase rotation affects the steady-state amplitude and phase balance of every single wave, and thus all six equations become coupled in a microresonator system. We note that the variation of the field amplitudes through a variation in the pump detuning characterizes the FM-to-AM conversion of the system. This FM-to-AM conversion is a property present in all microresonator systems because a pump frequency fluctuation changes the position of the pump relative to the resonance and thus affects the amount of power coupled in. In addition, as a result of the delicate amplitude and phase balance in Eq. (15), the AM-to-FM process also exists in SBS oscillators as we will see in Section VI.

In order to simplify our analysis, we assume the pump detuning to be zero $(\sigma_F = 0)$ and also assume operation at the SBS gain peak $(\Omega = \Omega_b)$ so that $\sigma_F, \sigma_B, \sigma_\rho = 0$ from Eq. (16). Under these conditions, the steady state of Eq. (15) requires that $\phi_F - \phi_B - \phi_\rho = \pi/2$ and $\phi_S = \phi_F$. With this set of assumptions, the amplitude and phase equations become effectively decoupled. For example, a frequency fluctuation of the pump still causes the phases of the individual waves to rotate. However, since $\phi_F - \phi_B - \phi_\rho$ is stabilized around the value of $\pi/2$, $\sin(\phi_F - \phi_B - \phi_\rho)$ is stabilized at the peak of sinusoid and is thus only affected to second order by phase fluctuations. Physically, this demonstrates that the FM-AM conversion is minimized at the peak of the resonator's Lorentzian transfer function, as one would expect.

We can further show that these assumptions also effectively decouple the phase equations from amplitude fluctuations. For example, if we introduce a fluctuation of the pump amplitude in $\partial\phi_F/\partial t$, we find that $\partial\phi_F/\partial t$ is affected only to second order through perturbations of amplitude × phase. This occurs because $\phi_S = \phi_F$ in steady state for our operating conditions specified earlier, and thus only pump amplitude fluctuations which occur concurrently with phase fluctuations affect $\partial\phi_F/\partial t$. Similar arguments can be made for fluctuations in the amplitude of the forward, backward, and density waves, which apply generally to the rest of the phase equations in Eq. (15).

Assuming $\sigma_F = 0, \Omega = \Omega_b$, we now proceed to perturb Eq. (15) in order to determine a set of linearized coupled-mode equations for small-signal perturbations of the SBS oscillator. We first analyze the case of amplitude fluctuations introducing $|A'_{B,F}| \to |A'_{B,F}| + \delta|A'_{B,F}|$, $|\rho'| \to |\rho'| + \delta|\rho'|$, and $|S'| \to |S'| + \delta|S'|$ into Eq. (15) and cancelling out the steady-state response. Here $\delta|A'_{B,F}|, \delta|\rho'|, \delta|S'|$ denote amplitude fluctuations of the backward/forward, density, and pump waves. Introducing these perturbations into Eq. (15) yields

$$\frac{\partial \delta |A'_F|}{\partial t} = -\frac{\delta |A'_F|}{2\tau_F} - \frac{\gamma_e \omega_\ell}{4n_0^2 \rho_0} \Lambda_F \left( |\rho'|\delta|A'_B| + |A'_B|\delta|\rho'| \right)$$

$$+ \sqrt{\frac{1}{\tau_{ext}}} \delta |S|$$

$$\frac{\partial \delta |A'_B|}{\partial t} = -\frac{\delta |A'_B|}{2\tau_B} + \frac{\gamma_e \omega_\ell}{4n_0^2 \rho_0} \Lambda_B \left( |\rho'|\delta|A'_F| + |A'_F|\delta|\rho'| \right)$$

$$\frac{\partial \delta |\rho'|}{\partial t} = -\frac{\Gamma_b}{2} \delta|\rho'| + \frac{\epsilon_0 \gamma_e \ell_\rho^2}{4\Omega R^2} \Lambda_\rho \left( |A'_F|\delta|A'_B| + |A'_B|\delta|A'_F| \right) + f_r$$

(23)

where we have introduced $f_r$ as a Langevin white Gaussian noise source describing the real component of fluctuations in the density wave [40]. This noise is thermally driven and serves as the force which first initiates the spontaneous Brillouin scattering process. However, once a coherent field is developed, the incoherence of the spontaneous process introduces fluctuations in both the phase and amplitude of the oscillation signal. Assuming this noise is equipartitioned into the real and imaginary quadratures, $\langle f_r(t) f_r^*(t') \rangle = C\delta(t-t')$ [40] where $C = kT\rho_0 \Gamma_b / v^2 V$ is the autocorrelation strength of $f_r(t)$, $k$ is Boltzmann's constant, and $V$ is the acoustic mode volume. Note that we have ignored the effects of shot noise in Eq. (23).

Because we have assumed the presence of only one mode within the SBS gain bandwidth in our derivation [see Eqs. (1), (2), and (6)], all of the generated gain and noise couples into the lone oscillating SBS wave. To account for multiple modes, one can represent $\tilde{E}_B$ in Eq. (2) as a superposition of all backward waves in the system and appropriately partition the nonlinear polarization generated by the forward and density waves [Eq. (6)] between the backward waves. Should the need arise, a similar technique can also be applied to account for multiple acoustic modes.

Our goal now is to solve Eq. (23) separately for a perturbation of the supplied pump and for a thermally-driven perturbation of the density wave [32]. In our calculations, we assume $\Lambda_F \approx \Lambda_B$ since the backward and forward waves are close in mode number. We analyze the case of a pump amplitude fluctuation first by converting to the frequency domain and applying Cramer's rule to Eq. (23). This analysis yields

$$\delta|A'_F|(\omega) = \frac{\delta|S|}{\sqrt{\tau_{ext}}} \frac{(j\omega/2)(1/\tau_B + \Gamma_b) - \omega^2}{\Delta_A}$$

$$\delta|A'_B|(\omega) = \frac{\delta|S|}{\sqrt{\tau_{ext}}} \frac{(|A'_B|/|A'_F|)(1/2\tau_B)(j\omega + \Gamma_b)}{\Delta_A} \quad (24)$$

$$\delta|\rho'|(\omega) = \frac{\delta|S|}{\sqrt{\tau_{ext}}} \frac{(|\rho'|/|A'_F|)(\Gamma_b/2)(j\omega + 1/\tau_B)}{\Delta_A}$$

with the determinant

$$\Delta_A = -j\omega^3 - \omega^2 \left( \frac{1}{2\tau_F} + \frac{1}{2\tau_B} + \frac{\Gamma_b}{2} \right)$$

$$+ j\omega \left( \frac{|A'_B|^2}{|A'_F|^2} \frac{1}{2\tau_B} + \frac{1}{2\tau_F} \right) \left( \frac{1}{2\tau_B} + \frac{\Gamma_b}{2} \right) + \frac{\Gamma_b}{2\tau_B^2} \frac{|A'_B|^2}{|A'_F|^2} \quad (25)$$

Here $\omega$ denotes the frequency at which the system response is evaluated. This response is third order due to the interaction of the forward, backward, and density waves. We note that the forward wave response has a zero in the numerator at $\omega = 0$, with no equivalent pole in the denominator to cancel this zero. This zero nulls the response of the forward wave at DC, as one would expect, because the forward wave amplitude becomes clamped by the steady-state operation of the SBS oscillator.

The denominator of Eq. (24) in principle provides information on the response of the system at resonance. However, analysis of Eq. (25) becomes difficult as the system behavior is third order. To proceed further, we assume that for our frequencies of interest, the term $-j\omega^3$ is negligible compared to the remaining imaginary component of Eq. (25). We will check the validity of this assumption at the end of our analysis. Ignoring the $-j\omega^3$ term, the system is now effectively second order with a resonance frequency of

$$\omega_{R_A}^2 = \frac{1}{\tau_B} \frac{|A'_B|^2}{|A'_F|^2} \times \frac{\Gamma_b}{2\tau_B} \left( \frac{1}{2\tau_F} + \frac{1}{2\tau_B} + \frac{\Gamma_b}{2} \right)^{-1} \quad (26)$$

In Eq. (26), we have explicitly separated the system resonance into the product of two components. The first component $\left[ |A'_B|^2 / \left( |A'_F|^2 \tau_B \right) \right]$ is the inverse of the forward wave stimulated lifetime and characterizes the decay time of the forward wave as it supplies gain for the stimulated Brillouin scattering process. This can be seen by attempting to write the second term on the right-hand side of the first equation in Eq. (18) as $-|A'_F|/(2\tau_{stim})$. Assuming $\Lambda_F \approx \Lambda_B$, substitution of the second equation in Eq. (18) at steady state into the first yields the desired expression for the stimulated lifetime $(\tau_{stim})$. The second component of Eq. (26) is effectively the lifetime of the slower wave (between the backward wave and the density wave). For example, if $1/\tau_B \gg 1/\tau_F, \Gamma_b$ the second component yields $\Gamma_b$. However, if $\Gamma_b \gg 1/\tau_{B,F}$, the second component in this case yields $1/\tau_B$. If the lifetime of the forward wave is fastest, the second component then becomes a mixture of the lifetimes of all three waves. Therefore, similar to the case of a semiconductor laser [32], $\omega_{R_A}^2$ of Eq. (26) is inversely proportional to the product of the stimulated emission lifetime and the photon/phonon lifetime.

From Eqs. (25) and (26), the damping ratio of this system can also be determined. To simplify our analysis, we assume that $\Gamma_b \gg 1/\tau_{B,F}$. This assumption generally holds true for high-Q microresonators at 1550 nm wavelength where $\Gamma_b/2\pi \sim 10$–100 MHz and $1/2\pi\tau_{B,F} < 1$ MHz. With these assumptions, we find the system damping ratio to be

$$\zeta_A = \left(\frac{1}{2\tau_B}\frac{|A'_B|^2}{|A'_F|^2} + \frac{1}{2\tau_F}\right) \bigg/ \left(\frac{2}{\tau_B}\frac{|A'_B|}{|A'_F|}\right) \qquad (27)$$

The numerator of Eq. (27) is inversely proportional to the total lifetime of the forward wave $(1/\tau_{stim} + 1/\tau_F)$, whereas the second term is twice the resonance frequency.

Our previous analysis can be interpreted by imagining the system consisting of the forward, backward, and density waves in their equilibrium state. At a given instance of time, the pump wave experiences a sinusoidal modulation of its amplitude perturbing the system from its steady-state operation. If the modulation is slow compared to the response time of the system, the forward wave remains approximately clamped at its steady-state value, while the backward and density waves experience a sinusoidal modulation in response to the pump [Eq. (24)].

If we increase the modulation frequency slowly, we find that at some frequency $\omega_{R_A}$, the response of the system is at resonance with the forward wave exchanging energy with the backward and density waves. That is, as the pump amplitude increases, the forward wave initially increases in response to the pump as according to Eq. (18). The increase of the forward wave increases the SBS gain, which then prompts the growth of the backward and density waves. As the backward and density waves increase further, the gain becomes depleted and begins to decrease. This causes the backward and density waves to decrease and restarts the cycle until the oscillations settle to steady state. If the system is perturbed at this resonance frequency, these oscillations are reinforced leading to the buildup of a resonance response. Furthermore, if we assume the density wave lifetime to be fast, then this process primarily involves the interplay of the forward and backward waves [the slower wave in Eq. (26)], as the density wave will respond as needed to the current state of the system.

Assuming the density wave response to be fast, we expect the damping of these oscillations to increase as the response times of the SBS gain and backward wave become further separated. For example, the resonance becomes weaker if the SBS gain can quickly respond to changes of the backward wave such that a prolonged ringing of the two waves does not occur. This can be seen from Eq. (27). If we assume the gain response $(\tau_{stim})$ to be fast, then we can neglect the second term in the first parenthesis of Eq. (27).

For this case, the damping ratio takes the form $\zeta_A = (\tau_B|A'_F|)/(4\tau_{stim}|A'_B|)$. Thus, to achieve the largest damping, one should reduce the system losses so that $\tau_B$ is maximized while driving the system with sufficient strength so that the stimulated SBS lifetimes are kept to a minimum.

Our previous analysis was based on the assumption that Eq. (25) could be modeled as a second-order system by ignoring the $-j\omega^3$ term. We now verify the validity of this assumption. We assume $\Gamma_b \gg 1/\tau_{B,F}$ and substitute Eq. (26) into Eq. (25) comparing the strength of the imaginary terms. Since $\Gamma_b\tau_B \gg 1$, we find that the $-j\omega^3$ term can be effectively ignored for a high-Q resonator. For resonators where the forward, backward, and density waves have similar lifetimes, the full system response of Eq. (25) must be used.

It is useful at this point to calculate the spectral densities of the fluctuating variables in Eq. (24) as spectral densities are ultimately what are determined through experimental measurement. These spectral densities can be found through multiplying $\delta|A'_F|(\omega), \delta|A'_B|(\omega), \delta|\rho'|(\omega)$ by $\delta|A'_F|(\omega'), \delta|A'_B|(\omega'), \delta|\rho'|(\omega')$, taking the time average, and integrating over $\omega'$ [32]. This operation yields

$$S^P_{\delta|A'_F|} = \frac{S^P_{\delta|S|}}{\tau_{ext}|\Delta_A|^2}\left[\omega^4 + \frac{\omega^2}{4}\left(\frac{1}{\tau_B} + \Gamma_b\right)^2\right]$$

$$S^P_{\delta|A'_B|} = \frac{S^P_{\delta|S|}}{\tau_{ext}|\Delta_A|^2}\frac{|A'_B|^2}{4|A'_F|^2\tau_B^2}(\omega^2 + \Gamma_b^2) \qquad (28)$$

$$S^P_{\delta|\rho'|} = \frac{S^P_{\delta|S|}}{\tau_{ext}|\Delta_A|^2}\frac{|\rho'|^2}{|A'_F|^2}\frac{\Gamma_b^2}{4}\left(\omega^2 + \frac{1}{\tau_B^2}\right)$$

where $S^P_{\delta|A'_F|}, S^P_{\delta|A'_B|}, S^P_{\delta|\rho'|}, S^P_{\delta|S|}$ denote the spectral densities of fluctuations for the forward, backward, density, and pump waves, respectively.

We now return to Eq. (23) and analyze the system response to a thermal excitation of the density wave. The effect of the thermal excitation is shown in Fig. 3 for a single occurrence of a noise event [13, 32]. Note that $f_r$ in Eq. (23) describes a continuous stream of noise perturbations in amplitude along with their associated rate of occurrence. In Fig. 3, a noise event with amplitude $|\rho_N|$ and relative phase $\theta_N$ results in a small-signal amplitude $(\Delta|\rho'|)$ and phase $(\Delta\phi_\rho)$ perturbation of the density wave. Since the phase of the noise event is random with respect to the coherent density wave, $\theta_N$ is uniformly distributed between 0 and $2\pi$. An amplitude perturbation of the density wave changes the amount of coupling between the forward and backward waves [see $\partial |A'_B|/\partial t$ in Eq. (15)] and thus causes a perturbation of the SBS amplitude. We quantify the cumulative effects of amplitude noise by solving for $\delta |A'_{B,F}|, \delta |\rho'|$ in Eq. (23).

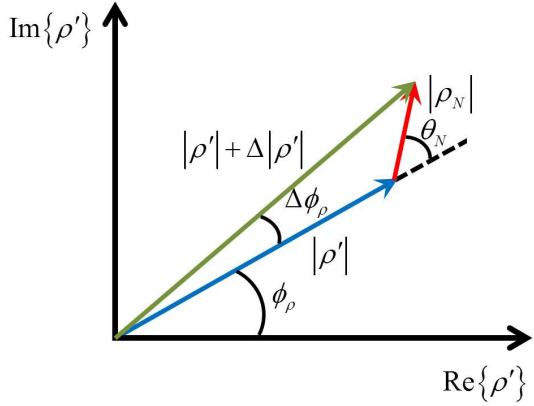

Fig. 3. Illustration of a density wave amplitude and phase perturbation due to a single noise event.

To do so, we convert Eq. (23) to the frequency domain and apply Cramer's rule [32] to find

$$\delta |A'_F|(\omega) = -\frac{f_r(\omega)}{\Delta_A} \frac{|A'_F|}{2\tau_{stim} |\rho'|} \left( j\omega + \frac{1}{\tau_B} \right)$$

$$\delta |A'_B|(\omega) = \frac{f_r(\omega)}{\Delta_A} \frac{|A'_B|}{2\tau_B |\rho'|} \left( j\omega + \frac{1}{2\tau_F} - \frac{1}{2\tau_{stim}} \right) \quad (29)$$

$$\delta |\rho'|(\omega) = \frac{f_r(\omega)}{\Delta_A} \left[ \begin{array}{c} -\omega^2 + j\omega \left( \frac{1}{2\tau_F} + \frac{1}{2\tau_B} \right) \\ + \frac{1}{2\tau_B} \left( \frac{1}{2\tau_F} + \frac{1}{2\tau_{stim}} \right) \end{array} \right]$$

where $1/\tau_{stim} = \left[ |A'_B|^2 / \left( |A'_F|^2 \tau_B \right) \right]$ as before. Setting $\omega = 0$ in Eq. (28), one can verify that the resulting perturbations satisfy the steady state of Eq. (23). Through a similar procedure to that of Eq. (28), we find the spectral densities of the perturbations to be

$$S^P_{\delta|A'_F|} = \frac{S^P_{f_r}}{|\Delta_A|^2} \frac{|A'_F|^2}{4\tau^2_{stim} |\rho'|^2} \left( \omega^2 + \frac{1}{\tau_B^2} \right)$$

$$S^P_{\delta|A'_B|} = \frac{S^P_{f_r}}{|\Delta_A|^2} \frac{|A'_B|^2}{4\tau^2_B |\rho'|^2} \left[ \omega^2 + \left( \frac{1}{2\tau_F} - \frac{1}{2\tau_{stim}} \right)^2 \right]$$

$$S^P_{\delta|\rho'|} = \frac{S^P_{f_r}}{|\Delta_A|^2} \left[ \omega^2 \left( \frac{1}{2\tau_F} + \frac{1}{2\tau_B} \right)^2 + \left( -\omega^2 + \frac{1}{2\tau_B} \left( \frac{1}{2\tau_F} + \frac{1}{2\tau_{stim}} \right) \right)^2 \right]$$

(30)

where $S^P_{f_r}$ is the spectral density for thermally-excited in-phase fluctuations of the density wave. Since the noise process considered here is white (memoryless), $S^P_{f_r}(\omega) = C$ over all frequencies where $C$ is the autocorrelation strength of $f_r(t)$ defined earlier. Note that Eqs. (28) and (30) differ by the noise source which initiates perturbations of the forward, backward, and density waves.

Comparing Eqs. (24) and (29), we find that unlike the case of a pump fluctuation in Eq. (24), the forward wave in the case of a density fluctuation is no longer completely clamped at $\omega = 0$. This occurs because the total increase in forward, backward, and density wave amplitude must individually balance their respective losses in the steady state. For the case of a density fluctuation, we observe from Eq. (23) that this additive fluctuation adds to/reduces the supplied gain provided by the forward wave depending on the sign of $f_r$. Thus for perturbations of the density wave, the forward wave amplitude must continuously take on different values so that the total increase in the density wave due to SBS amplification *and* noise compensates for the losses of the density wave.

Comparing Eqs. (28) and (30), we observe that the resonance response of the system governed by $|\Delta_A|^2$ is similar for fluctuations of the pump and density wave. However, accounting for the total response, we find that the overall system behavior to be dissimilar. For example, at high frequencies faster than the characteristic lifetimes of the system, we expect the forward wave to fall as $1/\omega^2$ for a pump perturbation and as $1/\omega^4$ for a thermal fluctuation of the density wave. On the other hand, the rolloff of the backward wave at high frequencies is $1/\omega^4$ for both cases.

### VI. SBS LASER FREQUENCY NOISE

We now return to Eq. (15) analyzing the SBS oscillator's phase response to noise. Careful attention should be paid to the distinction between phase and frequency in this analysis. The phase need not settle to a single value at steady state since a constant rotation in phase yields a frequency shift. We can develop intuition on the fundamental noise limit of the SBS laser by considering the process by which noise propagates to the generated SBS signal.

We begin by considering a thermally-induced white Gaussian Langevin noise source ($f$) that drives fluctuations of the density wave in Eq. (14). $f$ has autocorrelation strength $\langle f(t) f^*(t') \rangle = 2 \langle f_r(t) f_r^*(t') \rangle$, which accounts for the equipartition of noise energy between the real and imaginary quadratures. Its strength is determined by imparting $kT/2$ of noise energy to each degree of freedom of the acoustic mode under thermal equilibrium conditions [40]. As before, Fig. 3 shows the perturbation of the density wave amplitude and phase induced by a single noise event. The density wave interacts with the forward wave to generate a reflected wave that becomes added to the SBS signal (shown in-phase in Fig. 4). An amplitude fluctuation of the density wave directly results in an amplitude fluctuation of the SBS signal, as was found in Section V. A phase fluctuation of the density wave changes the relative phase of the superimposed waves, and thus affects the resulting phase of the backwards wave.

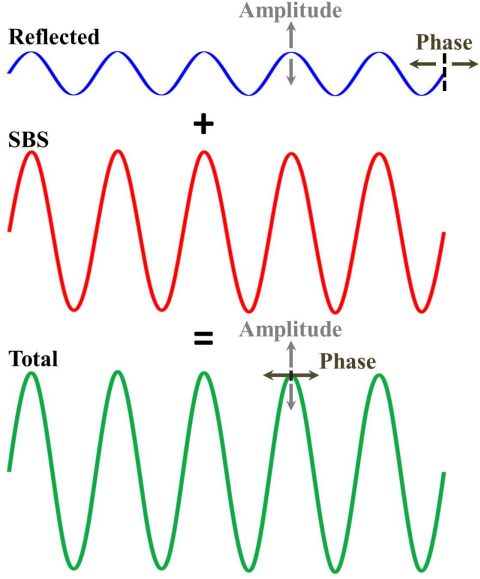

Fig. 4. Illustration of noise propagating into a perturbation of the total SBS wave.

With this intuition, we are now interested in determining the fundamental limits to the SBS laser noise. As before, we assume $\sigma_F = 0, \Omega = \Omega_b$ so that $\sigma_F, \sigma_B, \sigma_\rho = 0$, $\phi_F - \phi_B - \phi_\rho = \pi/2$, and $\phi_S = \phi_F$. These operating conditions also effectively decouple the evolution of phase from that of amplitude in Eq. (15). With these assumptions, the phase response of the system to small-signal density perturbations is given by

$$\frac{\partial \delta\phi_F}{\partial t} = -\frac{\gamma_e \omega_\ell}{4n_0^2 \rho_0} \frac{|\rho'||A'_B|}{|A'_F|} \Lambda_F \left( \delta\phi_\rho + \delta\phi_B - \delta\phi_F \right)$$
$$- \sqrt{\frac{1}{\tau_{ext}}} \frac{|S|}{|A'_F|} \delta\phi_F$$
$$\frac{\partial \delta\phi_B}{\partial t} = \frac{\gamma_e \omega_\ell}{4n_0^2 \rho_0} \frac{|\rho'||A'_F|}{|A'_B|} \Lambda_B \left( \delta\phi_F - \delta\phi_B - \delta\phi_\rho \right) \quad (31)$$
$$\frac{\partial \delta\phi_\rho}{\partial t} = \frac{\epsilon_0 \gamma_e \ell_\rho^2}{4\Omega R^2} \frac{|A'_F||A'_B|}{|\rho'|} \Lambda_\rho \left( \delta\phi_F - \delta\phi_B - \delta\phi_\rho \right) + \frac{f_{im}}{|\rho'|}$$

In Eq. (31), we have introduced the Langevin noise source $f_{im}$ which governs out-of-phase perturbations of the density wave. Its autocorrelation strength is equal to that of $f_r$, i.e. $\langle f_{im}(t) f_{im}^*(t') \rangle = C\delta(t-t')$, since the thermal noise excitation is equipartitioned between the real and imaginary quadratures. $\delta\phi_{B,\rho,F}$ represent small-signal phase perturbations of the backwards, density, and forward waves.

Converting Eq. (31) to the frequency domain and assuming $\Lambda_F \approx \Lambda_B$, we find

$$\delta\phi_F(\omega) = -\frac{f_{im}(\omega)}{|\rho'|\Delta_\phi} \frac{j\omega}{2\tau_{stim}}$$
$$\delta\phi_B(\omega) = -\frac{f_{im}(\omega)}{|\rho'|\Delta_\phi} \frac{1}{2\tau_B} \left( j\omega + \frac{1}{2\tau_F} + \frac{1}{2\tau_{stim}} \right) \quad (32)$$
$$\delta\phi_\rho(\omega) = \frac{f_{im}(\omega)}{|\rho'|\Delta_\phi} \left[ \begin{array}{c} -\omega^2 + j\omega\left(\frac{1}{2\tau_B} + \frac{1}{2\tau_F}\right) \\ + \left(\frac{1}{2\tau_F} + \frac{1}{2\tau_{stim}}\right) \frac{1}{2\tau_B} \end{array} \right]$$

where

$$\Delta_\phi = -j\omega^3 - \omega^2 \left( \frac{1}{2\tau_B} + \frac{\Gamma_b}{2} + \frac{1}{2\tau_F} \right)$$
$$+ j\omega \left( \frac{1}{2\tau_F} + \frac{1}{2\tau_{stim}} \right) \left( \frac{1}{2\tau_B} + \frac{\Gamma_b}{2} \right) \quad (33)$$

Note that we have simplified Eqs. (32) and (33) using $|S|/(|A'_F|\sqrt{\tau_{ext}}) = 1/2\tau_F + 1/2\tau_{stim}$, which can be derived from Eq. (19). Examining Eq. (32), we find that $\delta\phi_{B,\rho}$ diverges in the steady state ($\omega = 0$). This can be seen from Eq. (31) where no general solution exists when the time derivatives are set to zero. If we account for the Langevin noise source $f_{im}$ in $\partial\phi_\rho/\partial t$ of Eq. (15), we see that $f_{im}$ takes the role of an additional force driving phase rotations of the density wave. Similar to the case of the SBS gain peak detuning, the frequencies of the backwards and density waves must shift [see Eq. (16)] in order to provide the necessary counter phase rotation to satisfy steady state.

Multiplying Eq. (32) by $j\omega/2\pi$, we see that the left-hand side becomes a description of the frequency fluctuations driven by thermal perturbations of the density wave. The corresponding frequency noise spectra are given by

$$S^P_{\delta v_F} = \frac{1}{16\pi^2} \frac{S^P_{f_{im}}}{|\rho'|^2 |\Delta_\phi|^2} \frac{\omega^4}{\tau^2_{stim}}$$

$$S^P_{\delta v_B} = \frac{1}{16\pi^2} \frac{S^P_{f_{im}}}{|\rho'|^2 |\Delta_\phi|^2} \frac{1}{\tau_B^2}\left[\omega^4 + \omega^2\left(\frac{1}{2\tau_F} + \frac{1}{2\tau_{stim}}\right)^2\right] \quad (34)$$

$$S^P_{\delta v_\rho} = \frac{1}{4\pi^2} \frac{S^P_{f_{im}}}{|\rho'|^2 |\Delta_\phi|^2} \begin{bmatrix} \omega^6 + \omega^4\left(\frac{1}{2\tau_B}+\frac{1}{2\tau_F}\right)^2 \\ -\omega^4\left(\frac{1}{2\tau_{stim}}+\frac{1}{2\tau_F}\right)\frac{1}{\tau_B} \\ +\omega^2\left(\frac{1}{2\tau_{stim}}+\frac{1}{2\tau_F}\right)^2 \frac{1}{4\tau_B^2} \end{bmatrix}$$

In Eq. (34), $S^P_{\delta v_{B,\rho,F}}$ represent the frequency noise spectral densities associated with the backward, density, and forward waves, while $S^P_{f_{im}}$ denotes the spectral density of out-of-phase fluctuations due to thermal excitation of the density wave. Since the thermal noise process is white (memoryless), $S^P_{f_{im}}(\omega) = C$ over all frequencies.

We are interested in determining the fundamental noise level achievable by the SBS laser. The SBS frequency noise spectrum of Eq. (34) is white at lower Fourier frequencies and decays upon reaching larger Fourier frequencies. Setting $\omega = 0$ for $S^P_{\delta v_B}$, we find

$$S^P_{\delta v_B}(\omega = 0) = \frac{1}{4\pi^2} \frac{S^P_{f_{im}}}{|\rho'|^2 (\Gamma_b \tau_B + 1)^2} \quad (35)$$

The SBS white frequency noise level can be understood if we first assume $\Gamma_b \tau_B \gg 1$ (high-Q microresonator) and then substitute $|\rho'| = |A'_B|/2\tau_B g|A'_F|$ into Eq. (35). Here,

$$g = \frac{\gamma_e \omega_\ell}{4 n_0^2 \rho_0} \Lambda_B \quad (36)$$

is the SBS gain coefficient which when multiplied by $|\rho'||A'_F|/|A'_B|$ describes the transfer of phase fluctuations of the density wave to rotations of the SBS wave [see Eq. (31)]. The equality $|\rho'| = |A'_B|/2\tau_B g|A'_F|$ is derived from the steady state condition of $\partial |A'_B|/\partial t$ in Eq. (15) with $\phi_F - \phi_B - \phi_\rho = \pi/2$. With these substitutions in Eq. (35), we find

$$S^P_{\delta v_B}(\omega = 0) \approx \frac{1}{\pi^2} \frac{1}{|A'_B|^2} g^2 |A'_F|^2 \frac{S^P_{f_{im}}}{\Gamma_b^2} \quad (37)$$

Since $S^P_{f_{im}}(\omega) = C = \Gamma\langle\rho_N(t)\rho_N^*(t)\rangle/2$ [40] and because the integration over the noise spectrum of the density wave ($S^P_{\rho_N}$) yields its total power ($S^P_{\rho_N} \Gamma_b \approx \langle\rho_N(t)\rho_N^*(t)\rangle$), the term $S^P_{f_{im}}/\Gamma_b^2 = S^P_{\rho_N}/2$ represents the spectrum of out-of-phase fluctuations exhibited by the density wave induced by thermal noise. These density fluctuations interact with the forward wave ($g|A'_F|$) to couple noise into the generated SBS wave. Since the noise power is constant for a given forward wave amplitude (which is clamped at threshold), the resulting perturbation of the SBS wave decreases with increasing backward wave energy ($|A'_B|^2$). To minimize the white noise floor of the SBS signal, one should reduce the lasing threshold $|A'_F|^2$ thereby minimizing the noise coupled into the SBS wave. It is similarly important to also maximize the power of the backwards wave so that the perturbations introduced by the noise are comparatively small relative to the total signal level. We note that Eqs. (35) and (37) can be further modified to account for additional noise contributions beyond the fundamental SBS limit (e.g., thermorefractive noise [41, 42]).

As in the amplitude noise case of Section V, the coupling between the forward, backwards, and density waves creates a resonance in the system response before the spectrum of the noise decays. For fluctuations in phase [Eq. (32)], this response consists of a pole at $\omega = 0$. However, since we multiply by $j\omega/2\pi$ to determine the frequency noise response, this pole becomes cancelled revealing a white frequency noise floor. For frequency noise fluctuations, the system resonance occurs at

$$\omega^2_{R_v} = \left(\frac{1}{2\tau_F} + \frac{1}{2\tau_{stim}}\right)\left(\frac{1}{2\tau_B} + \frac{\Gamma_b}{2}\right) \quad (38)$$

with a damping ratio of

$$\zeta_v = \frac{\frac{1}{2\tau_B} + \frac{\Gamma_b}{2} + \frac{1}{2\tau_F}}{2\sqrt{\left(\frac{1}{2\tau_F}+\frac{1}{2\tau_{stim}}\right)\left(\frac{1}{2\tau_B}+\frac{\Gamma_b}{2}\right)}} \quad (39)$$

For high-Q cavities with $\Gamma_b/2 \gg 1/2\tau_{B,F}$, the damping ratio simplifies to

$$\zeta_v \approx \frac{1}{2}\sqrt{\Gamma_b / \left(\frac{1}{\tau_F} + \frac{1}{\tau_{stim}}\right)} \quad (40)$$

Therefore, for high-Q resonators, the system is typically damped except for cases when the system is driven with sufficient strength such that $1/\tau_{stim}$ approaches $\Gamma_b$.

## VII. SBS LASER STEADY-STATE SIMULATIONS

In the following sections, we use numerical techniques to simulate the complex behavior of the SBS laser. Our simulation is carried out iteratively using Eq. (14) with an additional Langevin noise source $f$ driving perturbations of the density wave. $f$ is thermally induced and has autocorrelation strength $\langle f(t) f^*(t') \rangle = 2C\delta(t-t')$ with $C = kT\rho_0 \Gamma_b / v^2 V$. Starting from noise, the evolution of the forward, backward, and density waves is tracked over time which allows for the determination of the laser's steady state and dynamic behavior. A listing of the parameters used in our simulation is provided in Table 1. For simplicity, we have taken the modal overlaps to be approximately 0.5 and have assumed operation at 1550 nm wavelength. The acoustic $(V)$ and optical $(V_{ph})$ mode volume is similarly assumed to be approximately equal corresponding to a modal area of 25 μm² and a resonator radius of 2.8 mm. The values of $\Gamma_b$ and $\Omega_b$ for the acoustic wave are derived from Ref. [33] for a silica resonator operated at 1550 nm. Finally, the values of $S$ and $V_{ph}$ in Table 1 correspond to a laser pump power of 1 mW.

Using the parameters of Table 1, we first determine the steady-state operation of the SBS laser as a function of the pump detuning. Typically the experimental operation of the microcavity SBS laser requires tuning the pump into the cavity resonance, and thus this simulation serves as a useful aid for analyzing the SBS laser's behavior as the pump frequency is swept. Figure 5(a) shows the normalized transmission past the microresonator for several values of pump power. The horizontal axis depicts the detuning of the pump from the cavity resonance in units of the resonator linewidth. At a pump power of 0.01 mW, the supplied power is below the SBS lasing threshold at every value of detuning, and thus the transmission traces out the cavity's characteristic Lorentzian interference pattern. Note that effects of self- and cross-phase modulation as well as thermal drift were not accounted for in our simulation, which would otherwise lead to asymmetry in the transmission profile.

Increasing the pump power to 0.25 mW, the SBS laser reaches threshold near a detuning of 1.1 × the resonator linewidth. Once the SBS reaches threshold, the forward wave power begins to clamp but can still change depending on the phase of the SBS gain [see Eq. (19)]. In addition, the phase variation of the forward wave with detuning also diverges from that of a conventional resonator in order to satisfy steady state with the backward and density waves. Since $T = A'_F / \sqrt{\tau_{ext}} - S$, these properties alter the transmission past the microresonator so that the profile is no longer a pure Lorentzian. In particular, the clamping of the forward wave power results in incomplete cancellation of the pump with the outcoupled forward wave causing the normalized transmission of Fig. 5(a) to flatten. This effect can be clearly observed at higher pump powers where the SBS lasing is seen to turn on at larger cavity detunings with the consequence of a diminished dip in transmission.

Figure 5(b) illustrates the outcoupled SBS power for a pump power of 1 mW. The SBS reaches threshold at a pump detuning of 2.1 linewidths offset from its resonance peak, as can also be inferred from Fig. 5(a). The outcoupled SBS power reaches a maximum of 0.285 mW when operated with zero detuning. The corresponding normalized intracavity forward wave power can be observed in Fig. 5(c). The forward wave power initially builds up as the pump detuning approaches zero. However, once the SBS oscillation reaches threshold, the forward wave power begins to clamp with slow variations depending on the phase of the SBS gain. For our parameters in Table 1, the SBS gain is maximum at zero pump detuning, and thus the required forward wave power for lasing is minimized at this point [see Fig. 5(c)]. Figure 5(d) illustrates the SBS gain $(\phi_F - \phi_B - \phi_\rho)$ as a function of the pump detuning for a pump power of 1 mW. Since the SBS and density waves are incoherent until oscillation is reached, we have restricted the range of pump detuning in Fig. 5(d) from -2 to 2 cavity linewidths, corresponding to the points where SBS lasing occurs. As predicted from our analysis, the phase of the SBS

Table 1. Parameters used for the simulation of the SBS laser

| Parameter | Simulation Value |
|---|---|
| $\tau_F$ | $1/(2\pi \times 2 \times 10^6)$ s |
| $\tau_B$ | $1/(2\pi \times 2 \times 10^6)$ s |
| $\tau_{ext}$ | $1/(2\pi \times 1.33 \times 10^6)$ s |
| $\Lambda_F$ | 0.5 |
| $\Lambda_B$ | 0.5 |
| $\Lambda_\rho$ | 0.5 |
| $\gamma_e$ | 1.5 [33, 43] |
| $\omega_\ell$ | $1.22 \times 10^{15}$ rad/s |
| $n_0$ | 1.5 |
| $\rho_0$ | 2200 kg/m³ |
| $S$ | $1.52 \times 10^{10}$ $\sqrt{W/m^2 F}$ |
| $\Gamma_b$ | $2\pi \times 15.64 \times 10^6$ rad/s [33] |
| $\Omega_b$ | $2\pi \times 11.55 \times 10^9$ rad/s [33] |
| $\Omega$ | $2\pi \times 11.55 \times 10^9$ rad/s |
| $\ell_\rho / R$ | $1.22 \times 10^7$ 1/m |
| $V$ | $4.33 \times 10^{-13}$ 1/m³ |
| $V_{ph}$ | $4.33 \times 10^{-13}$ 1/m³ |

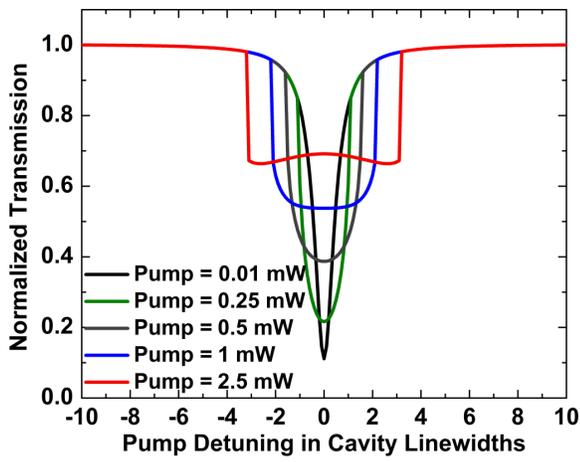

(a)

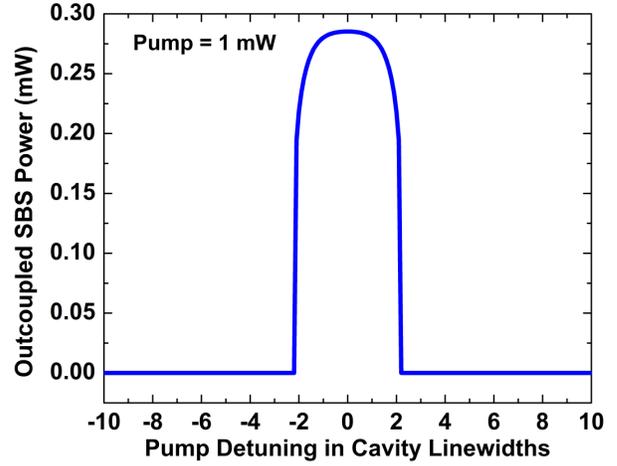

(b)

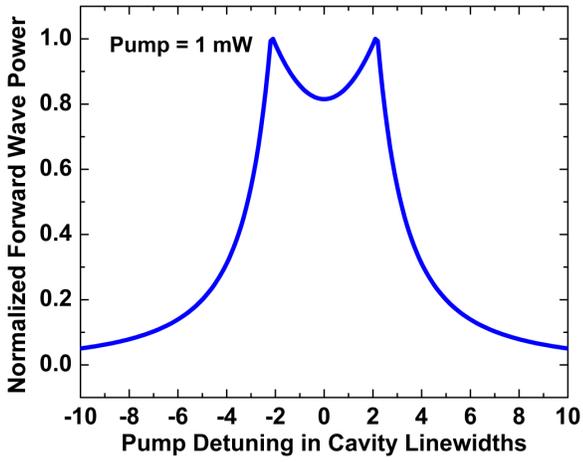

(a)

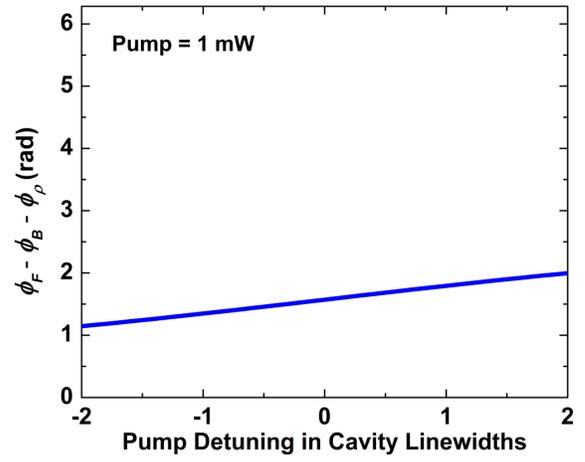

(b)

Fig. 5. Simulated SBS laser (a) normalized transmission past the cavity, (b) outcoupled power, (c) normalized intracavity forward wave power, and (d) gain phase as a function of pump detuning (in units of cavity linewidths). The simulations of (b), (c), and (d) correspond to a pump power of 1 mW.

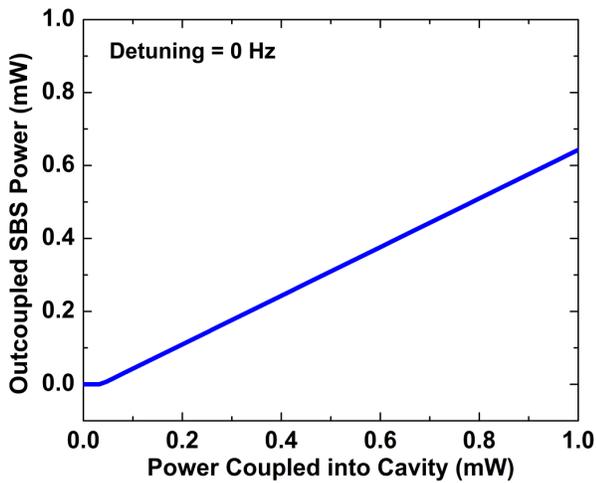

Fig. 6. SBS laser outcoupled power as a function of the total input power supplied into the cavity.

gain is $\pi/2$ at zero detuning and slowly rotates from this value when operated off of the SBS gain peak.

The previous simulations characterize the SBS laser operation as a function of cavity detuning for a fixed pump power of 1 mW. We now set the detuning to be zero and investigate the laser's performance when the pump power is varied. We are primarily interested in the amount of SBS power obtainable for a given supply of power into the microresonator cavity. Figure 6 shows the outcoupled SBS power versus the total amount of power coupled into the cavity and thus depicts the input-output relationship of the laser [Eq. (22)]. The simulated lasing threshold occurs at 0.032 mW with a corresponding slope efficiency of 66.7 %, which agrees with the analytical expression of Eq. (22) using the values provided in Table 1. However, we note that although the slope efficiency of the coupled-in power is 66.7 %, there is an additional efficiency loss from the supplied pump power that does not couple into the cavity

[see Fig. 5(a)]. For example, a total pump power of 4.1 mW was required to couple 1 mW of power into the cavity thus yielding a coupling efficiency of 24.4 %.

### VIII. SIMULATIONS OF FUNDAMENTAL NOISE LIMIT

We are now interested in analyzing the fundamental limits of noise achievable by the SBS laser. As before, the simulation is seeded with white Gaussian noise of random phase, which provides the initial kick for self-oscillation. By continuously seeding the microresonator with noise, the phase and amplitude of the forward, backward, and density waves become perturbed from their steady-state values. Since $\partial \phi_{B,\rho,F}/\partial t = 0$ at steady state, the frequency noise of these corresponding waves is directly found from simulating their phase evolution in Eq. (15) with noise introduced into the system. Figure 7(a) illustrates the simulated and analytical SBS frequency noise spectrum for zero pump detuning and a pump power of 1 mW. The corresponding outcoupled SBS power is 0.29 mW. Note that since measured values of frequency noise are often defined through a single-sided spectrum, we have doubled our calculated SBS frequency noise, which effectively maps the negative frequencies onto the positive frequencies. The SBS frequency noise is white at lower offset frequencies and exhibits a resonance at ~6 MHz. Beyond ~10 MHz, the frequency noise response rolls off at -20 dB/decade. Note that since the effects of shot noise have not been accounted for, the simulated rolloff continues on for higher offset frequencies.

From Fig. 7 (a), the simulated white frequency noise floor is 0.47 Hz²/Hz at lower offset frequencies, which matches the analytical value for the SBS laser's white frequency noise of 0.51 Hz²/Hz [Eq. (35)]. These values also agree well with experimental measurements of the SBS laser's white frequency-noise floor [27, 44]. Since the linewidth and frequency noise of a laser are closely related, we can achieve an estimate of the SBS laser's linewidth using Fig. 7(a). If we approximate the entire frequency noise spectrum to be white, the corresponding power spectral density exhibits a Lorentzian lineshape with a full-width half maximum linewidth of 1.5 Hz. The simulated linewidth is 3 to 6 orders of magnitude narrower than conventional semiconductor or fiber lasers and thus highlights the excellent noise properties exhibited by the SBS gain medium.

The SBS laser's amplitude noise can also be simulated by analyzing the system's response to noise. The relevant figure of merit here is the laser's RIN, which provides a measure of the laser's amplitude fluctuations normalized to its signal level. The laser RIN can be related to the amplitude noise spectral density [Eq. (30)] by noting that $\left(P_O + \delta P_O\right) \propto \left(|A'_B| + \delta|A'_B|\right)^2$ and thus $\delta P_O \propto 2|A'_B|\delta|A'_B|$. The first expression is a statement that the SBS intracavity optical power $(P_O)$ and its associated noise fluctuation $(\delta P_O)$ are proportional to the square of the total intracavity SBS field amplitude (signal + noise). The second expression identifies the dominant contribution to the power fluctuation, which consists of a heterodyne between the signal amplitude and noise amplitude fluctuation. Since a laser's RIN is measured as a ratio of noise power to signal power in the electrical domain after photodetection, we convert $\delta P_O \propto 2|A'_B|\delta|A'_B|$ into its corresponding power spectral density and normalize with respect to $P_O^2 \propto |A'_B|^4$. The factors of proportionality all cancel, and we thus find

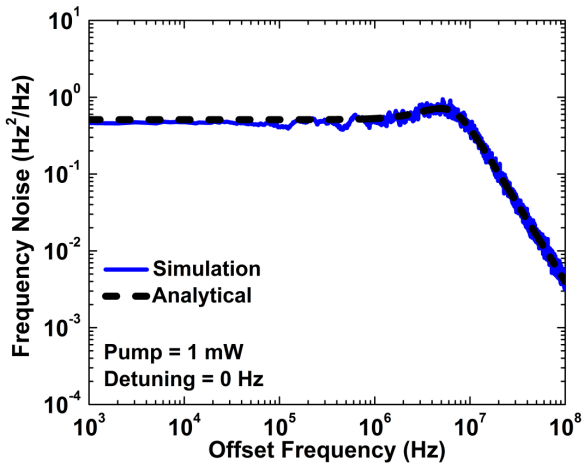 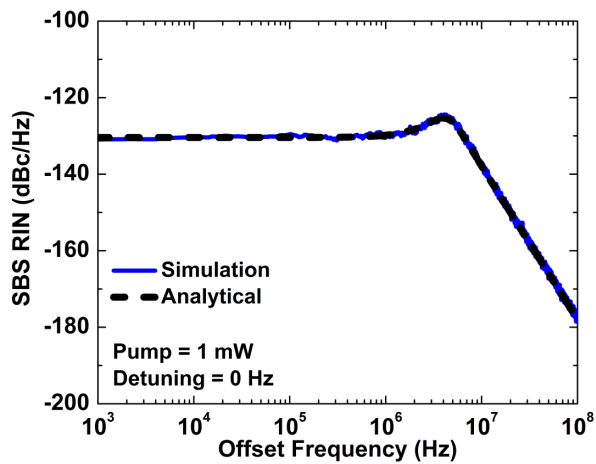

(a)        (b)

Fig. 7. Simulated (blue line) and analytical (black dashed line) SBS laser (a) frequency noise and (b) RIN for a pump power of 1 mW and pump detuning of 0 Hz.

$$\text{RIN} = \frac{8 S^P_{\delta|A'_B|}}{|A'_B|^2} \quad (41)$$

Note that since laser RIN is defined as single-sided, we have introduced an additional factor of two in Eq. (41) which maps the negative frequencies onto the positive frequencies. Depending on the noise source (pump fluctuation or thermally-induced density wave fluctuation), either Eq. (28) or Eq. (30) may be used for calculating RIN. Our analytical calculations of the intrinsic SBS laser amplitude noise here are performed using Eq. (30) in Eq. (41).

Figure 7b shows the simulated and analytical SBS RIN for a pump power of 1 mW and a pump detuning of 0 Hz. The relaxation oscillation resonance occurs near ~4 MHz, which matches the approximate resonance frequency of 3.5 MHz [Eq. (26)]. The SBS RIN is white at a level of -130.4 dBc/Hz at lower offset frequencies and decays at -40 dB/decade past the relaxation resonance.

From Fig. 7, we observe that although the SBS laser exhibits excellent performance in frequency noise, its RIN characteristics appear worse compared to lasers of semiconductor or fiber technology. One reason for this degradation in laser RIN is due to the lower optical powers exhibited by the SBS laser. From Eq. (30), we see that $S^P_{\delta|A'_B|}$ scales inversely with $|A'_B|^4$ $(\propto P_O^2)$ when the stimulated lifetimes are slower than the intrinsic cavity decay rate. Note that the density wave scales proportionally to the backward wave [see Eq. (17)] as the two waves mutually promote each other's growth. The normalization of Eq. (41) introduces an additional factor of $|A'_B|^2$ which yields a combined cubic inverse dependence of RIN on SBS power. If $\tau_{stim} \ll \tau_F$, the total RIN scaling instead becomes first order inversely with SBS power. With an increase of outcoupled SBS power from 0.29 mW to 29 mW, the SBS laser RIN can be reduced by at least 20 dB. However, these larger optical powers cannot usually be achieved experimentally in SBS microresonator lasers since the SBS wave eventually becomes the pump for the next Stokes order [not modelled in Eq. (14)].

In addition to the limitations due to optical power, the effects of oscillator feedback also have a relatively minor impact on the SBS laser's amplitude fluctuations. In any oscillator, the gain is stabilized to the value that compensates the total system loss, thereby reducing the oscillator's exhibited amplitude noise. For example in a laser, an increase of the intracavity power causes saturation of the gain below the intracavity loss, which then results in attenuation of the power until steady state is reached. A similar self-stabilization is observed in Eq. (23) for the SBS laser. We see that if the density wave is instantaneously increased by means of a noise fluctuation $(\delta|\rho'| > 0)$, the system develops a driving force to increase the SBS wave due to the SBS interaction of the forward wave with the density wave. However, this process acts to deplete the power in the forward wave $(\delta|A'_F| < 0)$, which then reduces the available SBS gain. These two processes can in principle balance one another resulting in complete cancellation of the SBS wave's amplitude noise. This effect can be more concretely observed in Eq. (29) where for simplicity we consider the laser's operation at zero frequency. As per Eq. (23), we multiply $\delta|A'_F|$ by $|\rho'|$ and $\delta|\rho'|$ by $|A'_F|$ and subsequently sum them together. It is clear from this operation that the fluctuations of the forward and density waves can exactly compensate one another when $\tau_F = \tau_{stim}$. This can also be verified by setting $\omega = 0$ for $\delta|A'_B|$ in Eq. (29) along with setting $\tau_F = \tau_{stim}$. However, since $\tau_{stim}$ is dependent on the operating point, the typical operation of the SBS laser results in the fluctuations of the forward wave under- or over-compensating the fluctuations of the density wave.

## IX. SIMULATIONS OF PUMP NOISE TRANSFER

In this section, we simulate the transfer of noise from the pump laser to the SBS wave. This noise increases the SBS laser noise above the intrinsic limits found in the previous section. To simulate the pump noise transfer, we apply a coherent small-signal modulation to either the amplitude or phase of the pump wave. We then divide the resulting frequency noise/RIN imprinted onto the SBS wave by the frequency noise/ RIN of the pump. For performing these simulations, the thermally-induced fluctuations of the density wave are switched off.

Figure 8(a) shows the transfer of pump frequency fluctuations to the SBS laser's frequency noise for a pump power of 1 mW and a pump detuning of 0 Hz. At lower offset frequencies, 1.3 % of the pump noise converts to fluctuations of the SBS wave. At higher frequencies, the noise reaches a damped resonance at ~5 MHz and then rolls off at -40 dB/decade when the cavity can no longer respond. Note that although only a small fraction of the pump frequency noise transfers over to the SBS laser, the transfer of pump noise to the density wave is nearly 100 %. The amount of noise transfer is governed by the response rate for each of the individual waves ($1/2\tau_B$ for the SBS wave and $\Gamma_b/2$ for the density wave) to phase perturbations. These phase response rates can be understood by substituting the steady state of the amplitude evolution equations [Eq. (15)] into Eq. (31). When the phase of the pump undergoes an instantaneous change, the phases of both the backward and density waves rotate to reach steady state. Although there is only one value of combined phase shift that satisfies steady state, no restrictions apply to the individual values of $\phi_B, \phi_\rho$. Thus, one desires the density wave response to be fast so that it absorbs the entirety of the phase rotation. To estimate the noise transfer into the SBS wave, we take the ratio of $1/2\tau_B$ to $\Gamma_b/2$ and square the result for conversion to spectral density. This operation yields 0.016 which agrees well with our simulated value of 0.013 at lower offset frequencies.

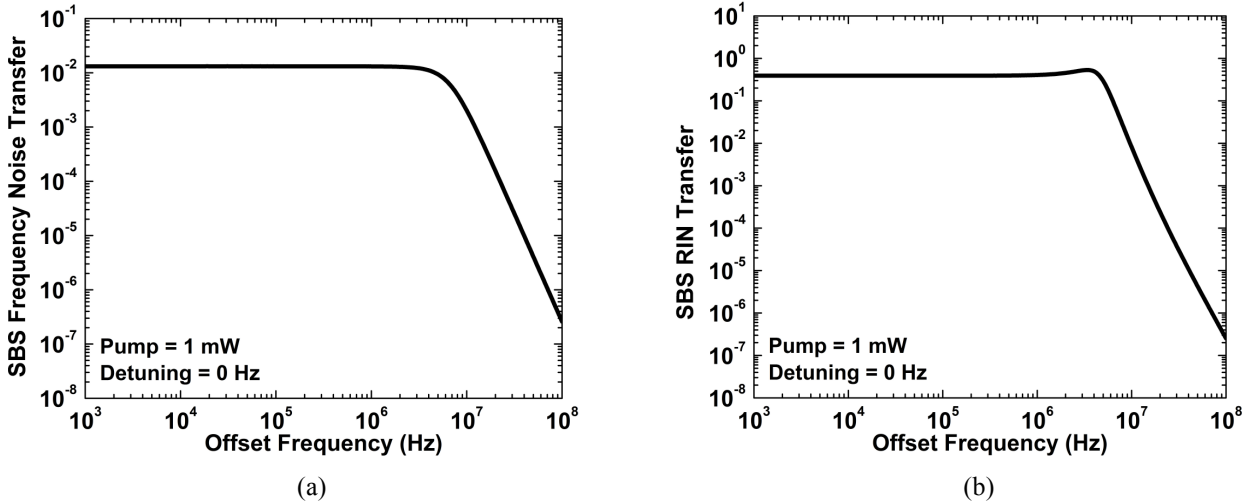

Fig. 8. Simulated pump to SBS laser (a) frequency noise and (b) RIN transfer for a pump power of 1 mW and zero pump detuning.

Figure 8(b) shows the corresponding conversion of pump RIN into SBS laser RIN. The RIN transfer is 39.1 % at lower offset frequencies which agrees with the analytically calculated value of 39.1 %. At ~4 MHz, the RIN transfer exhibits a resonance and then afterwards decays at even higher offset frequencies. Since the pump RIN is typically low, the RIN of the SBS laser is usually limited by its intrinsic amplitude fluctuations [see Fig. 7(b)].

Because of the delicate balance between phase and amplitude in a microresonator cavity, there will also be effects of noise conversion from amplitude to phase and from phase to amplitude. However, these processes are approximately zero when the microresonator is operated with zero pump detuning. Our simulations of these noise processes appear to vary depending on the simulation parameters. However, we consistently find the conversion of pump RIN to SBS frequency noise (pump frequency noise to SBS RIN) to be below the level of $10^{-25}$ Hz$^2$ ($10^{-23}$ 1/Hz$^2$) at a 10 kHz offset frequency.

In order to accurately assess the conversion of pump noise into the opposite quadrature, we increase the amount of detuning used in our simulations. Figure 9(a) shows the transfer of pump frequency noise into SBS RIN for a pump detuning of $1/\tau_B$ (one cavity linewidth) and a SBS gain detuning of $\Gamma_b$ (one SBS gain linewidth). To achieve enough SBS gain to self-oscillate, we increase the pump power to 2 mW. At this operating point, the out-coupled SBS power corresponds to 0.21 mW. From Fig. 9(a), the conversion of pump frequency noise into SBS RIN is $2.2\times10^{-13}$ 1/Hz$^2$ at lower offset frequencies. This number is intrinsically small as the level of RIN is much smaller than that of frequency noise for a typical laser. However, we see that the noise transfer at 10 kHz is much larger than that found with zero pump detuning ($< 10^{-23}$ 1/Hz$^2$). Beyond the system resonance near 3.5 MHz, the conversion of pump frequency noise to SBS RIN rolls off when the laser can no longer respond to pump fluctuations.

Figure 9(b) shows the conversion of pump RIN into SBS frequency noise again for the same operating conditions. The transfer response exhibits a steady increase of 20 dB/decade at lower offset frequencies thus indicating the presence of a low-frequency zero. Since frequency and phase are related by a derivative, the corresponding phase noise spectrum would be constant at low frequencies. A constant phase noise is intuitive as a fixed shift of the pump amplitude would result in the SBS laser settling into a different fixed steady-state phase arrangement. Since a shift in pump amplitude results in a constant shift in SBS phase at DC, the corresponding frequency fluctuation is zero, as can also be extrapolated from Fig. 9(b). For an offset frequency of 10 kHz, the noise transfer is $5.1\times10^7$ Hz$^2$ which is again significantly larger than the conversion noise under zero detuning ($< 10^{-25}$ Hz$^2$). Beyond the system resonance near 3.5 MHz, the system response initially rolls off at -60 dB/decade before changing slopes to -40 dB/decade past 16 MHz (approximately inverse of the phonon lifetime).

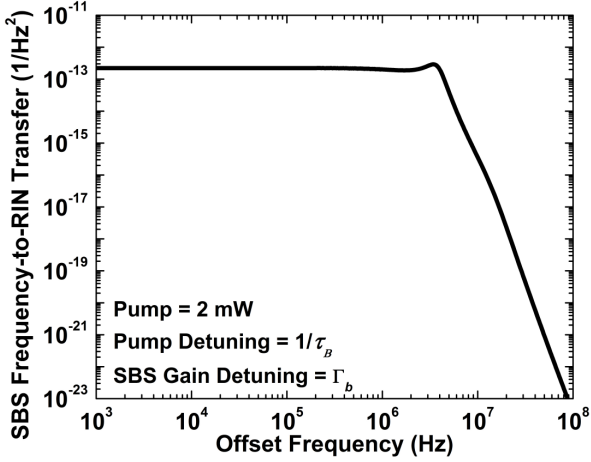
(a)

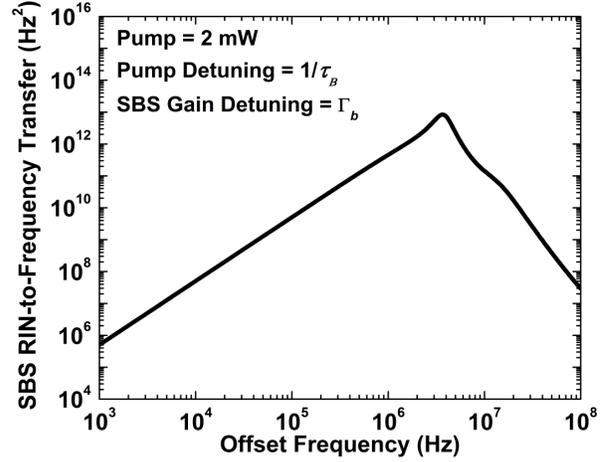
(b)

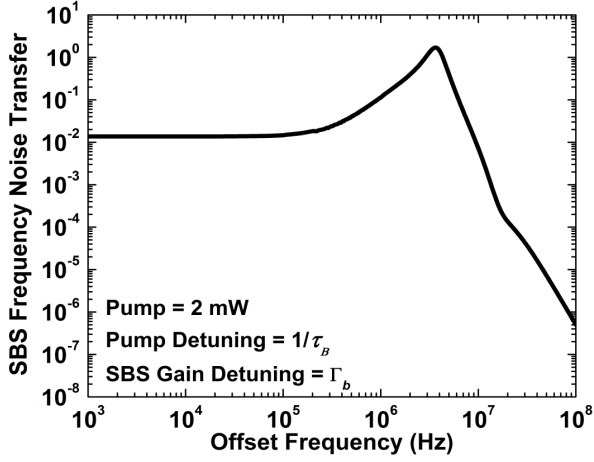
(c)

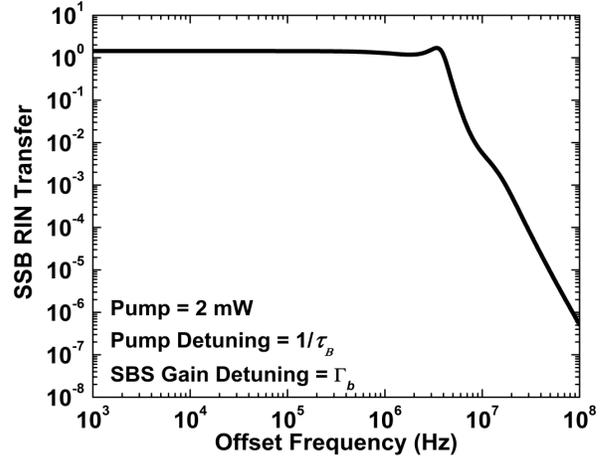
(d)

Fig. 9. Simulated (a) pump frequency noise to SBS RIN, (b) pump RIN to SBS frequency noise, (c) pump frequency noise to SBS frequency noise, and (d) pump RIN to SBS RIN transfer for a pump power of 2 mW, a pump detuning of $1/\tau_B$ (1 cavity linewidth), and a SBS gain detuning of $\Gamma_b$ (one SBS gain linewidth).

For completeness, Fig. 9(c) shows the transfer of pump frequency noise to the SBS laser's frequency noise when the pump detuning is $1/\tau_B$ and the SBS gain detuning is $\Gamma_b$. Similar to the zero detuning case [Fig 8(a)], 1.4 % of the pump frequency noise transfers over to the SBS wave at lower offset frequencies. The system response exhibits a sharp resonance near 3.5 MHz due to the strong coupling between amplitude and phase when the detuning is no longer zero. Beyond the system resonance, the rolloff is initially -60 dB/decade but changes to -40 dB/decade past 16 MHz.

Finally, Fig. 9(d) shows the transfer of pump RIN to the SBS laser's RIN for a pump detuning of $1/\tau_B$ and a SBS gain detuning of $\Gamma_b$. At low offset frequencies, the RIN transfer is 1.44× the RIN of the pump, which is 3.7× larger than that found with zero detuning [Fig 8(b)]. At higher frequencies, the noise transfer reaches a resonance near 3.5 MHz before finally stabilizing to a roll off of -40 dB/decade beyond 16 MHz. The properties of SBS noise transfer in Fig. 9 all closely match those found in experimental measurements [44].

## X. CONCLUSIONS

We have developed a set of coupled-mode equations that accurately describe the steady-state behavior and noise dynamics of the SBS laser. The coupling between the forward, backward, and density waves results in a complex noise response to amplitude or phase perturbation. Nevertheless, our analytical calculations and simulations show the potential for oscillation with hertz-class linewidths or below, enabled by the noise properties of the SBS gain. The intrinsic limits of SBS laser noise become degraded by a noisy pump due to the transfer of pump noise into the SBS wave. However, these effects are mitigated with the use of microcavities with higher Q. Our model can be readily extended to account for multiple oscillating modes, self- and cross-phase modulation nonlinearity, thermorefractive noise, or thermal bistability.

## XI. ACKNOWLEDGMENTS

We thank Dr. Frank Quinlan, Dr. Aurélien Coillet, and Prof. Kerry Vahala for their comments on this manuscript. This work was funded by NIST and the DARPA Pulse program. WL acknowledges support from the NRC/NAS. This work is a contribution of the US Government and is not subject to copyright in the US.